\documentclass[twocolumn,twocolappendix]{aastex6}

\usepackage{lipsum}
\usepackage{amsmath}
\usepackage{longtable}

\begin{document}


\title{Advection of potential temperature in the atmosphere of irradiated exoplanets: a robust mechanism to explain radius inflation}



\author{ 
P. Tremblin    \altaffilmark{1,2} and
G. Chabrier    \altaffilmark{2,3} and
N. J. Mayne    \altaffilmark{2}   and
D. S. Amundsen \altaffilmark{4}   and 
I. Baraffe     \altaffilmark{2,3} and
F. Debras      \altaffilmark{2,3} and
B. Drummond    \altaffilmark{2}   and
J. Manners     \altaffilmark{2,5} and
S. Fromang     \altaffilmark{6}
       }

\altaffiltext{1}{
  Maison de la Simulation, CEA-CNRS-UPS-UVSQ, USR 3441, CEA Paris-Saclay, 91191 Gif-Sur-Yvette, France} 

\altaffiltext{2}{
  Astrophysics Group, University of Exeter, EX4 4QL Exeter, UK}
  
\altaffiltext{3}{
  Ecole Normale Sup\'erieure de Lyon, CRAL, UMR CNRS 5574, 69364 Lyon
  Cedex 07, France}

\altaffiltext{4}{
  Department of Applied Physics and Applied Mathematics, Columbia
  University, New York, NY 10025, USA 
}

\altaffiltext{5}{
  Met Office, Exeter, EX1 3PB
}

\altaffiltext{6}{
  Laboratoire AIM, CEA/DSM-CNRS-Universit\'e Paris 7, Irfu/Service d'Astrophysique, CEA Paris-Saclay, 91191 Gif-sur-Yvette, France}

\email{pascal.tremblin@cea.fr}

\begin{abstract}
  The anomalously large radii of strongly irradiated exoplanets have remained a major puzzle in astronomy. Based on a 2D steady state atmospheric circulation model, the validity of which is assessed by comparison to 3D calculations, we reveal a new mechanism, namely the advection of the potential temperature due to mass and longitudinal momentum conservation, a process occuring in the Earth's atmosphere or oceans. At depth, the vanishing heating flux forces the atmospheric structure to converge to a hotter adiabat than the one obtained with 1D calculations, implying a larger radius for the planet. Not only do the calculations reproduce the observed radius of HD209458b, but also the observed correlation between radius inflation and irradiation for transiting planets. Vertical advection of potential temperature induced by non uniform atmospheric heating thus provides a robust mechanism explaining the inflated radii of irradiated hot Jupiters. 
\end{abstract}


\section{Introduction} \label{sec:intro}

The anomalously large radii of irradiated hot Jupiters is one of the most intriguing problems in our understanding of extrasolar giant planets. While various physical mechanisms have been proposed to resolve this puzzle \citep[see][for reviews]{Baraffe:2010fz,Fortney:2010jt,Baraffe:2014ha}, none of them provide, thus far, a satisfactory explanation. Generally speaking, they either lack a description of a robust mechanism  \citep[e.g. to explain downward transport of kinetic energy deep enough to reach the internal adiabat][]{Showman:2002ez}, or they must invoke fine-tuned conditions resistivity \citep[ohmic resistivity][enhanced opacities \citealt{Burrows:2007ca} or ongoing layered convection \citealt{Chabrier:2007ch}]{Batygin:2011ig}. On the other hand, there is now growing observational evidence to suggest a correlation between inflated radii and incident stellar flux \citep{Laughlin:2011hb,Miller:2011bq,Demory:2011ha,Weiss:2013kh,Figueira:2014hw} and the recently observed re-inflated hot Jupiter EPIC211351816.01 \citep{Grunblatt:2016tt} also supports this correlation (even though other explanations for this object might be possible). In the meantime, a large effort in the community has been dedicated to the development of general circulation models (GCMs) \citep[e.g.][see \citealt{Showman:2010to} for a review]{Showman:2009kl,DobbsDixon:2012jf,Mayne:2014cw}, yet so far none of these models have been able to reveal the long sought mechanism responsible for the inflated radii of hot Jupiters.
In this paper, we propose a new approach to the problem of the atmospheric circulation of tidally locked exoplanets. Starting from our 1D radiative/convective atmosphere code \texttt{ATMO} \citep{Amundsen:2014df,Amundsen:2017dq,Tremblin:2015fx,Tremblin:2016hi}, recently applied to irradiated giant planets \citep{Drummond:2016jg}, we construct a stationary 2D circulation model of the equatorial region of the planet (see Sect.~\ref{sec:eq} and Sect.~\ref{sec:boundary}). The originality (and key) of this approach is to look for a stationary solution while all existing 2D/3D atmospheric studies of irradiated hot Jupiters have only considered time-dependent simulations that cannot be run long enough to reach a steady state in the deep atmosphere, where the thermal timescale becomes prohibitevely long. In Sect.~\ref{sec:2d3d}, we show that the resulting pressure/temperature and wind structures from the 2D solution compare very well with the equatorial solution from the 3D simulation. In the deep atmosphere, the 2D solution naturally converges to an adiabatic structure before becoming unstable to convection and thus connects to a hotter adiabat than that obtained using standard 1D models, which, in turn, corresponds to a significantly larger radius for the planet. We derive a reduced model to explain this behavior, providing a consistent solution for the inflated radii of hot Jupiters: the circulation induces a vertical mass flux in the deep atmosphere, advecting the potential temperature in a region of small heating rate, thus imposing a hot adiabatic interior for the planet (see Sect.~\ref{sec:solution}). In Sect.~\ref{sec:correlation}, we show that the models reproduce the observed trend of increasing radii with increasing irradiation, and we provide conclusions and perspectives in Sect.~\ref{sec:conclusion}.

\section{Two-dimensional (2D) stationary circulation model} \label{sec:eq} 

To describe the atmosphere, we look for stationary solutions of the Euler equations in spherical coordinates, $r$ (radius), $\phi$ (latitude), and $\lambda$ (longitude) \citep{Mayne:2014cw}, where the velocity components are denoted by $u_r$, $u_\phi$  and  $u_\lambda$, density is $\rho$, pressure $P$, $\Omega$ the angular velocity and $g(r)$ the gravitational acceleration.  A heating rate $H_\mathrm{rad}$ is included, consistently calculated from the radiative transfer equation \citep{Amundsen:2016hw} (see the Appendix). We then look for a solution with $u_\phi=0$ (because of the north-south hemispheric symmetry) but with $\partial u_\phi/\partial \phi \neq 0$ and $u_\lambda,u_r \neq 0$ at the equator $\phi = 0$. 

The meridional momentum conservation condition trivially vanishes at the equator, so we lack an equation to find the longitudinal (zonal), vertical velocities, and the derivative of the latitudinal (meridional) velocity at the equator. The wind is essentially driven by the longitudinal pressure gradients, which implies gradients of longitudinal velocities through the longitudinal momentum conservation condition. In a steady state, mass conservation implies that the resulting longitudinal mass flow is balanced by a combination of the vertical and meridional ones. We assume that the geometry of the wind is defined by a constant form factor $\alpha$, which corresponds to the ratio of the meridional over vertical mass fluxes. This yields the extra equation: 

\begin{equation}\label{eq:alpha}
\frac{1}{r^2}\frac{\partial r^2 \rho u_r}{\partial r} = \frac{1}{r\alpha} \frac{\partial
  \rho u_\phi}{\partial \phi}  
\end{equation}

The physical significance of $\alpha$ can be highlighted by considering the following two limits. Namely, $\alpha \rightarrow \infty$ for a purely longitudinal and meridional wind and $\alpha\rightarrow 0$ for a purely longitudinal and vertical wind, respectively. We use the 3D GCM simulations of \citet{Amundsen:2016hw} to calibrate the profile of $\alpha$. Note that this factor is likely to depend on the gravity of the object. By injecting the mass conservation into the internal energy equation (see Appendix) we get, after some algebraic manipulation, the equation of entropy conservation written as a function of the advection of the potential temperature (defined as $\theta=T(P_0/P)^{(\gamma-1)/\gamma}$, where $P_0$  is the reference pressure, and $\gamma=C_p/C_v$, the ratio of the specific heats at constant pressure and volume, respectively):

\begin{equation}\label{eq:energy}
  -\frac{\gamma}{\gamma-1}P \vec{u}\cdot\vec{\nabla}(\ln\theta)=H_\mathrm{rad}
\end{equation}

The full system of equations can be rewritten in a more compact form (see Appendix). When there is no flow (i.e. zero velocity), the system of equations reduces to the equation of hydrostatic balance $\partial P/\partial r = -\rho g$ and a constant flux transport $F_\mathrm{rad}=\mathrm{cst}$, corresponding to our 1D scheme (in \texttt{ATMO}) in the absence of convection. We thus extend the same solver to include the three equations needed to compute $u_\lambda$, $u_r$, and $\partial u_\phi/\partial \phi$ at the equator, and we replace the constant flux transport equation of the 1D scheme by energy conservation including a heating rate (i.e. the derivative of the flux transport equation; see Appendix for details). 

\section{Boundary conditions and input parameters}\label{sec:boundary} 

The system of equations is solved on a fixed height grid between $R_\mathrm{min}$ and $R_\mathrm{max}$. The boundary conditions at the bottom of the atmosphere are (i) a linear extrapolation of the velocities from inside the computational domain (i.e. we keep a constant gradient at the boundary), (ii) an imposed maximum pressure and (iii) an imposed radiative flux $F_\mathrm{rad}=F_\mathrm{int}$, where $F_\mathrm{int}$ is the same as in the 3D calculation. At the top of the atmosphere, we use a linear extrapolation of all the variables ($\log(P)$ for the pressure). For an axisymmetric irradiation (or absence of irradiation), an
axisymmetric solution exists with $\partial u_\phi / \partial\phi =
0$, $u_r=0$, $u_\lambda =\mathrm{cst}$  and any 1D solution that obeys
hydrostatic balance and  constant flux transport conditions. Thus, the system as proposed is not able to fully specify the wind solution since any constant zonal wind is a solution with isotropic irradiation. This is not surprising as in 3D models the magnitude of the zonal wind is known to depend on the full system and on the dissipation processes (physical and numerical). Calculation of the magnitude of the zonal wind is a very active field of research with 3D GCMs and the determination of the zonal wind strength is still a matter of debate \citep[e.g.][]{Fromang:2016ji}.

\begin{figure}[t]
\centering
\includegraphics[width=0.98\linewidth]{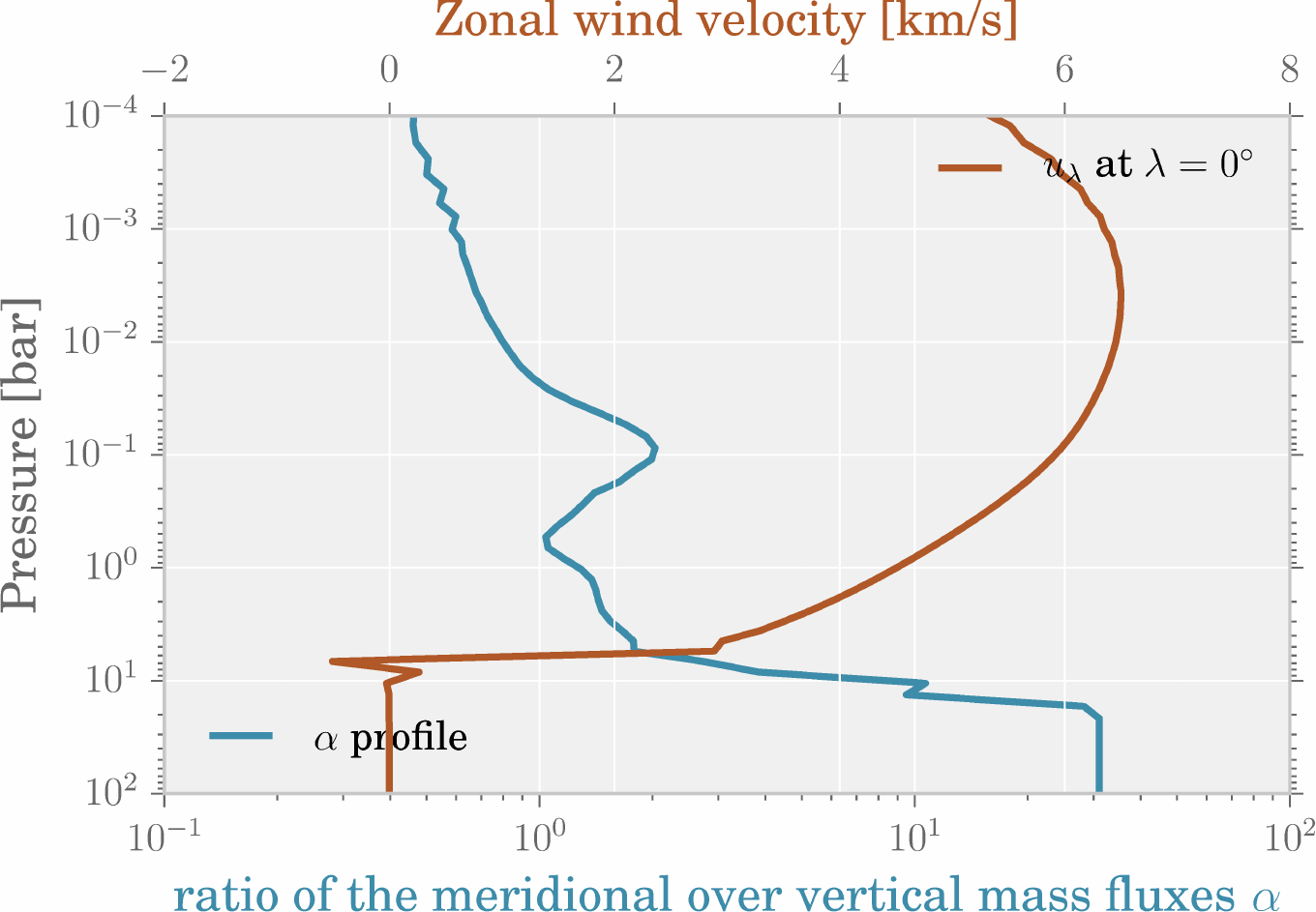}
\caption{\label{fig:inputs} Input profiles for the 2D stationary circulation model. The first input is the zonal wind profile at the substellar point ($\lambda$=0), and the second input is the $\alpha$ profile, defined as the ratio of the meridional to vertical mass flux at the equator.}
\end{figure}

In our 2D approach, we opt not to try to constrain the zonal wind magnitude but to parametrize it, a different but complementary approach to those trying to explain the equatorial superrotation \citep[e.g.][]{Vallis:2006kt,Showman:2011cb}. We impose a zonal-wind vertical profile at a given longitude of the equator (we choose the substellar point) and the 2D circulation model computes the full equatorial solution starting from this imposed profile. Therefore, the 2D circulation model depends on two inputs: a zonal wind profile at the substellar point and an $\alpha$ profile (the ratio of meridional to vertical mass flux) for the entire equatorial region. We choose input parameters that are derived from the 3D models to test our 2D solution against the equatorial solution of a 3D simulation;  the profiles are displayed in Fig.~\ref{fig:inputs}.
In order to determine $\alpha$, we decided to interpolate at the equator the values of the quantities of Eq.~\ref{eq:alpha} from the 3D GCM run of HD209458b from \citet{Amundsen:2016hw}. Then, at each vertical layer, we chose $\alpha$ such that it minimised the residual of Eq.~\ref{eq:alpha} with a least squares method on the longitude. The zonally averaged residual is smaller than 40\% of the zonally average vertical mass flux; hence we conclude that this approximation is relatively good. 
      The data from the 3D GCM were obtained after 1600 earth days of simulation, where the atmosphere has only converged above 1 bar. The values of $\alpha$ are nearly constant with time above that limit (we checked from 1000 to 1600 days), but are still evolving at deeper levels. In the region where the $\alpha$ value has not converged (for pressures larger than $\sim$20 bars), we decided to choose a constant value. We have tested $\alpha$=1, 30, 100 and found that the result is independent of this value.
The other usual input parameters include the gravity at the surface, the internal heat flux, the elemental composition, and the stellar and orbital properties, all given in Tab.~\ref{tab:param}.

\begin{table}
  \centering
  \begin{tabular}{l||r}
    Parameters & Values \\
    \hline
    \hline
    Vertical resolution & 50 \\
    Longitudinal res. & 10 \\
    P$_\mathrm{max}$ at R$_\mathrm{min}$ & 200 bar\\
    R$_\mathrm{min}$ & 1.287 R$_\mathrm{jup}$\\
    R$_\mathrm{max}$ & 1.48 R$_\mathrm{jup}$\\
    T$_\mathrm{int}$ & 100 kelvin\\
    logg & 2.97 \\
    $\gamma$ & 1.377 \\
    $\mu$ & 2.34 \\
    $\Omega$ & $2\times 10^{-5}$ s$^{-1}$\\
    F$_*$ Star spectrum & HD209458 (Kurucz)\\
    R$_*$ & 1.118 R$_\odot$ \\
    a     & 0.0475 au
  \end{tabular}
  \caption{List of the input parameters used in the 2D circulation model.}
  \label{tab:param}
\end{table}

\begin{figure*}[t]
\centering
\includegraphics[width=0.98\linewidth]{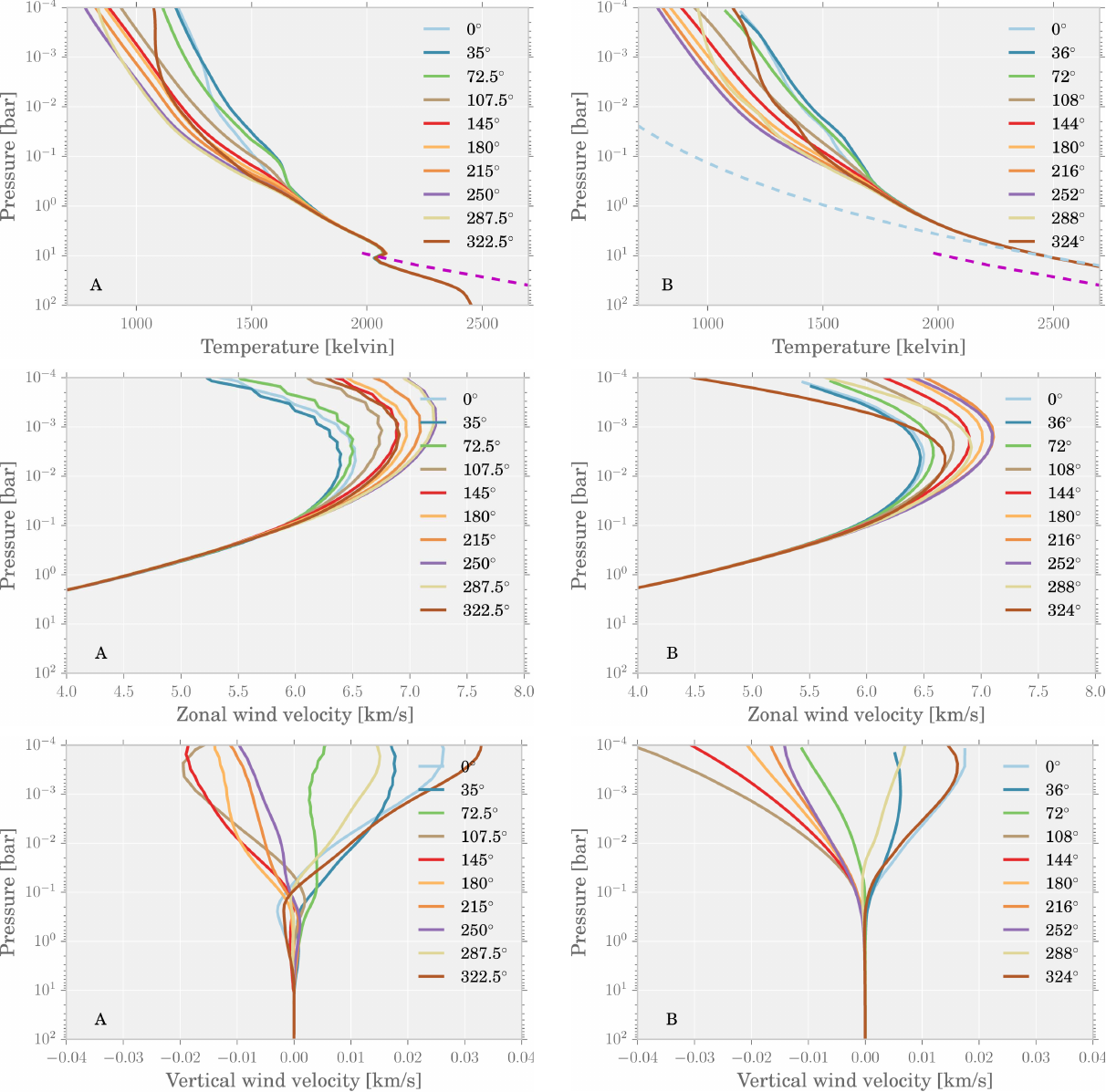}
\caption{\label{fig:profiles} Temperature, zonal wind, and vertical wind profiles as a function of pressure for different longitudes $\lambda$. (A): values extracted from a 3D run of the UM (21), (B): 2D stationary circulation model. The pink dashed line in the P/T plots portrays the thermal profile obtained in the evolutionary models with an extra-energy of 1\% of the stellar luminosity giving a radius of 1.4 $R_\mathrm{jup}$ \citep{Chabrier:2004db}, while the dashed cyan line in the P/T 2D plot corresponds to the deep atmosphere adiabat all profiles converge to ($\sim$1.6 $R_\mathrm{jup}$).}
\end{figure*}

\section{2D equatorial model versus 3D equator calculation}\label{sec:2d3d} 

\begin{figure}[t]
\centering
\includegraphics[width=0.98\linewidth]{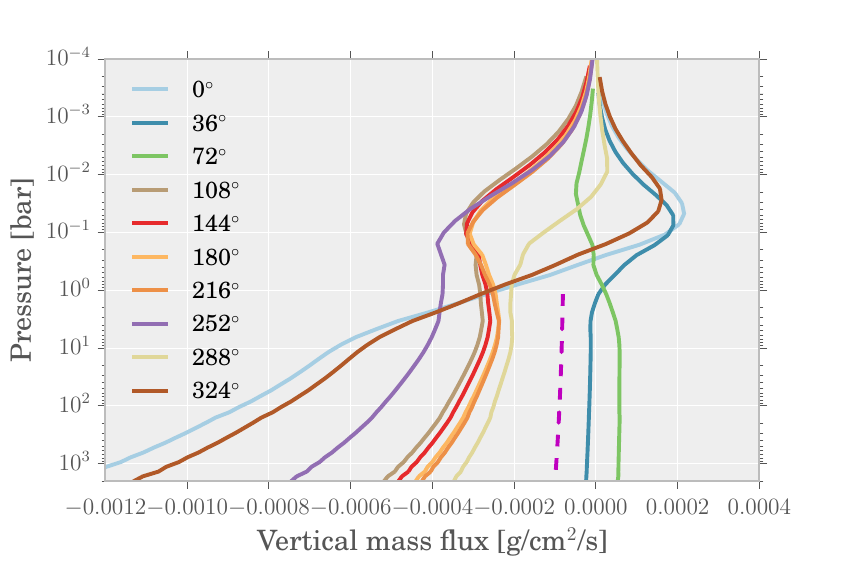}
\caption{\label{fig:massflux} Vertical mass fluxes as a function of pressure for different longitudes at the equator in the 2D steady-state model. The dashed magenta line represents the scaling of the mass fluxes over inverse squared radius when vertical motions are dominant in the mass conservation equation.}
\end{figure}

In this section, we compare the result from our 2D circulation model with those from the equatorial region of our 3D GCM simulations of HD209458b \citep{Amundsen:2016hw}. The results are portrayed in Fig.~\ref{fig:profiles} (A) and (B) respectively. A number of characteristics are well reproduced by the 2D solution, namely:
\begin{itemize}
\item The day-night temperature gradient
\item The shift of the hot spot in the upper atmosphere
\item The pressure at which the horizontal temperature gradients vanish
\item The variations of the longitudinal wind velocity
\item The global structure and magnitude of the vertical wind
\end{itemize}
Slight differences in the uppermost part of the atmosphere ($P$ $\leq$ 10$^{—3}$ bar) are likely caused by the different boundary treatments between the dynamical model and the 2D stationary model. The main differences between the 2D and 3D results are: (i) the pressure/temperature structure of the deep atmosphere (P $\geq$ 10 bar) and (ii) the differences between the vertical \text velocities in the deep atmosphere (mainly downwards between 1 and 10 bar). 
These differences stem from the fact that the 3D time-dependent solution has not yet converged towards a steady state solution. The 3D system tries to heat the deep interior from the initial 1D profile. This leads to artificial high vertical velocities in the deep atmosphere. Even these velocities, however, are unable to heat up the deep layers because of the extremely long thermal relaxation timescale at such depths. The system then tries to increase its energy and the internal heat flow is too small to efficiently heat up the deep layers on the timescale of the 3D simulation. This is an artificial problem, due to the fact that the initial profile is too cold. In reality, these objects loose energy by cooling, so it should be much easier to reach a steady state in the 3D runs by initializing with a hotter PT profile and let the system loose the extra energy \citep{Amundsen:2016hw}.  Even though the vertical velocities in the 2D deep atmosphere are small, however, the vertical mass flux remains significant, as shown in Fig.~\ref{fig:massflux}, and this behavior has important consequences, as explained in the next section. 

\section{Solution to the inflated-radii problem}\label{sec:solution} 

As illustrated by the dashed cyan line in Fig.~\ref{fig:profiles} (B), the 2D solution converges towards an adiabatic profile at significantly smaller pressures than the deeper inner convectively unstable regions (convection is not included in this 2D solution), which starts at P$\sim$40 bars for HD209458b with the proper 1.4 $R_\mathrm{Jup}$ radius \citep[see Fig. 1 of][]{Chabrier:2004db}. At these levels ($\sim$10 bars), the atmospheric and interior adiabats connect naturally, leading now to a hotter characteristic entropy profile for the planet than the one found in standard irradiated models. This in turn implies a larger radius for a given mass, $\geq$ 1.4 $R_\mathrm{jup}$ for HD209458b. The radius of 1.4 $R_\mathrm{jup}$ corresponds to the adiabat illustrated by the pink dashed line in the P/T plots, which is the one obtained in the evolutionary calculations including an extra energy equal to 1\% of the stellar luminosity in the planet thermal budget \citep{Chabrier:2004db} and the dashed blue line corresponds to a radius of 1.6 $R_\mathrm{jup}$. This solves the radius inflation problem: 2D atmospheric circulation leads to an adiabatic structure in the deep atmosphere before convection sets in, leading to a hotter internal adiabat. This behavior can be explained with the simple following model.
In deep enough layers (deeper than $\sim$1 bar), when the gradients of longitudinal velocities are negligible, the system of equations reduces to 

\begin{eqnarray}
  \frac{\partial r^2 \rho u_r}{\partial r} &=& 0 \cr
\frac{\partial p}{\partial \lambda}&=&0 \cr
\frac{\partial p}{\partial r} &\approx &- \rho g\cr
-u_r P\frac{\gamma}{\gamma-1}\frac{\partial \ln(\theta)}{\partial r} &=& H_\mathrm{rad} 
\end{eqnarray}

which can be rewriten as:
\begin{eqnarray}
r^2\rho u_r &=& R_0^2 \rho_0 u_{r,0}=\dot{M}_r\cr
\frac{\partial\ln\theta}{\partial \ln P }& =& \frac{\gamma-1}{\gamma}\frac{H_\mathrm{rad} r^2}{\dot{M}_r g}
\end{eqnarray}

Therefore, in the deep layers, as soon as the heating rate becomes sufficiently small, $H_\mathrm{rad} \ll \dot{M}_r g\gamma/(r^2 (\gamma-1))$, $\theta=\mathrm{cst}$, and thus $P\propto T^{\gamma/(\gamma-1)}$, which implies that the profile becomes adiabatic, a direct consequence of the vertical mass flux and the vertical transport of the potential temperature, even in the absence of convection. In turn, layers at pressure $P$ will be at temperature $T_0 (P/P_0 )^{(\gamma-1)/\gamma}$. From the 2D models, we get an estimate $H_\mathrm{rad} r^2 (\gamma-1)/\dot{M}_r g\gamma\approx 10^{-2}$ between 10 and 100 bars. We do not show the deepest layers in Fig.~\ref{fig:profiles} to focus on the comparison with the 3D run; however our results confirm that the profiles follow the adiabat down to $\sim$2 kbars. Furthermore, we show in the Fig.~\ref{fig:massflux} that the vertical mass flux follows the $1/r^2$ behaviour for $P$ $\geq$ 1 bar below the hot spot ($\lambda$= 36$^\circ$ and 72$^\circ$), as expected from mass conservation, confirming the validity of the reduced model at these longitudes. For other longitudes, the PT profile is maintained on the same adiabat because of the longitudinal advection of the potential temperature (see Eq.~\ref{eq:energy}). 
In the absence of radiatively driven vertical velocities, convection sets in at $P$ $\geq$ 40 bars \citep{Chabrier:2004db} but in the presence of circulation, the transition between the vertical global circulation and the convective zone might take place at deeper levels. In any case, our atmospheric adiabatic profile reconnects with a hotter internal adiabat than the conventional one. 
At this stage, it is worth stressing that this process differs entirely from the one proposed by \citet{Showman:2002ez}. Indeed, it is not related to the vertical transport of kinetic energy, a process hard to realize in a strongly gravitationally stratified medium, but to the vertical advection of the potential temperature (i.e. of entropy) due to a vertical mass flux and a small heating rate. The adiabatic adjustment is similar to the one proposed in \citet{Youdin:2010kp} with turbulent mixing, but in our 2D solutions the adjustment is simply induced by the vertical velocities of the global circulation itself. 
This process is very similar to the one taking place in the Earth stratified atmosphere or in deep oceanic layers in case of strong surface irradiation. The large-scale non uniform heating induces longitudinal motions (by conservation of longitudinal momentum) which in turn produce large air (in the atmosphere) or water (beneath the sea surface) mass fluxes which transport adiabatically the potential temperature to deep layers and thereby warm them up \citep[see chapter 16 in][and chapter 9.15 in \citealt{Gill:2016wx}]{Vallis:2006kt}.

We have verified that this process and the reduced system are independent of the value of $\alpha$ in the deep layers. A test with $\alpha=\infty$ deeper than 1 bar and $\alpha=2$ in the upper layers yields the same conclusions. Hence, our results are independent of the degree of convergence of the deep atmosphere in the 3D model. In the upper layers, the magnitude of the vertical wind depends directly on the value of $\alpha$ used at these levels. Hence $\alpha$ is also directly constrained by the high-altitude vertical wind of the 3D model, which is well converged for $P$ $\leq$ 10$^{-1}$ bar. Only when $\alpha$ is high everywhere in the atmosphere do the vertical velocities vanish, and we get back to the standard flat isothermal profile in the deep atmosphere (see Appendix). 

\begin{figure}[t]
\centering
\includegraphics[width=0.98\linewidth]{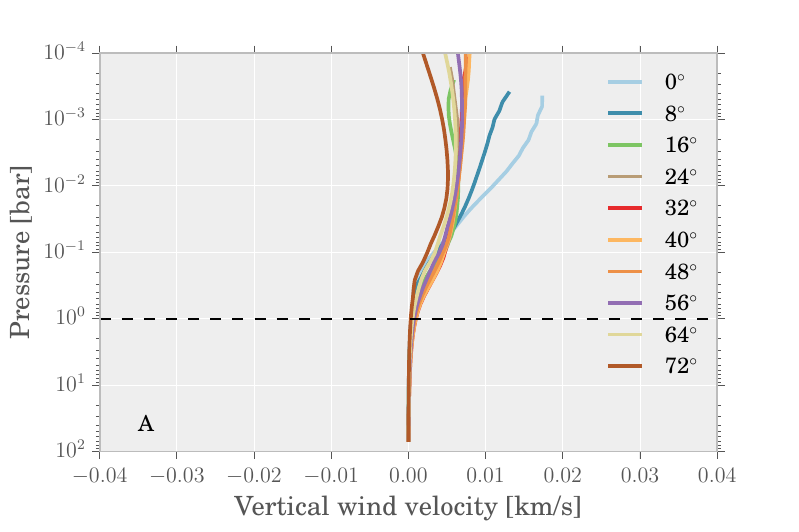}
\includegraphics[width=0.98\linewidth]{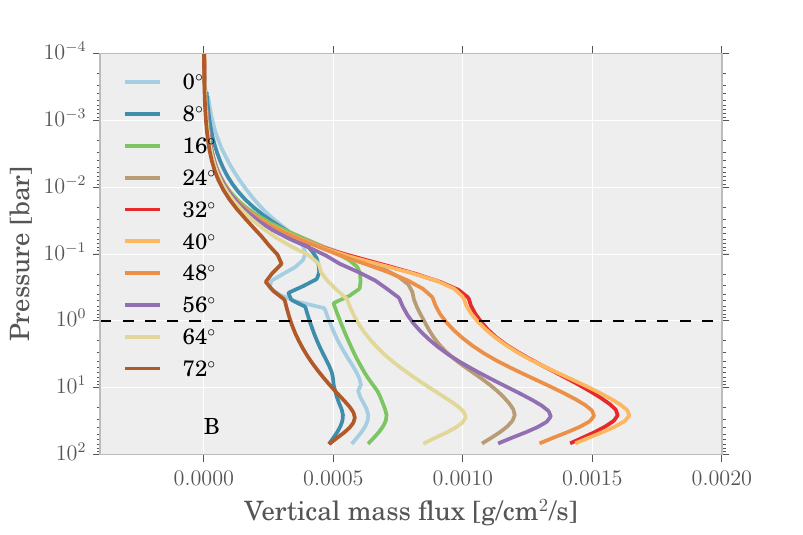}
\caption{\label{fig:massflux2} (A) For pressures less than 1 bar, the vertical wind profiles are extracted at different latitudes between 0$^\circ$ and 72$^\circ$ of the 3D UM simulation (at the longitude of the hot spot $\lambda=72^\circ$). At higher pressures the steady-state vertical velocities are approximated by integration of the vertical mass conservation. (B) shows the corresponding vertical mass fluxes.}
\end{figure}

The 2D model applies only to the equator of the planet. Since the longitudinal advection of the potential temperature maintains the 2D PT structure on the adiabat reached under the hot spot, we can expect that the meridional advection of potential temperature (see Eq.~\ref{eq:energy}) maintains the 3D PT structure on the same adiabat. Other latitudes of the full 3D solution in the UM simulation indeed seem to converge to the same adiabat as the 2D model, most likely because of the deep circulation identified in previous simulations \citep{Mayne:2014cw,Heng:2011dl}. Since the vertical mass flux is mainly downwelling in our equatorial solution, we expect the full 3D steady-state to produce upwelling flux at higher latitudes to globally conserve mass. We illustrate this possibility in Fig.~\ref{fig:massflux2} by displaying the upward vertical velocities at the high latitudes around the hot spot and by computing the deep vertical velocities with the vertical mass conservation condition. Current computational capabilities, however, prevent the possibility for 3D models to probe the long timescales needed for the circulation to reach this steady state down to the convection zone. The combination of both 3D runs and our 2D stationary circulation model thus enables us to better understand the short and long term dynamics of hot jupiters. It demonstrates that irradiation induced atmospheric circulation explains the inflated radii of hot jupiters.

Since the 2D solution reaches high temperatures in the deep atmosphere, the electric conductivity is high and ohmic dissipation might affect our results. We have estimated the efficiency of ohmic dissipation with Eq.~17 of \citet{Batygin:2011ig} and we get an efficiency of the order of 10$^{-5}$ below $\sim$8 bars. The reason is that even though the electric conductivity is high, the wind solution from the 3D simulation has almost zero velocity for $P$ $\geq$ 8 bars, hence no kinetic energy to be dissipated. As a consequence, ohmic dissipation should not impact our result as also found in other studies \citep{Rogers:2014ho,Rogers:2014iy}, even though future (lacking) constraints on the magnitude of the deep circulation in the atmosphere will help quantifying precisely this effect.

\section{Correlation of the inflated radii with increasing irradiation flux}\label{sec:correlation} 

\begin{figure}[t]
\centering
\includegraphics[width=0.98\linewidth]{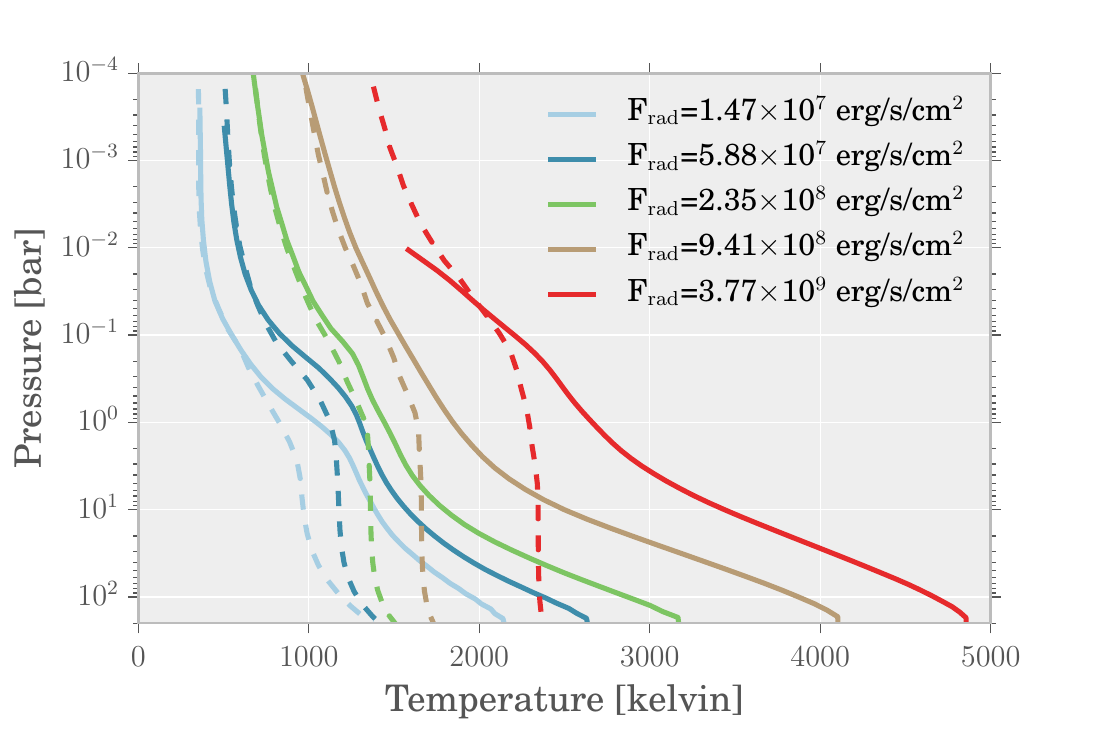}
\caption{\label{fig:pt_profiles} Average PT profiles of the 2D circulation model as a function of irradiation flux (solid lines); the standard irradiation for HD209 is $9.41\times10^8$ erg/s/cm$^2$. For comparison, we show the PT profiles obtained in 1D in dashed lines.}
\end{figure}

\begin{figure}[t]
\centering
\includegraphics[width=0.98\linewidth]{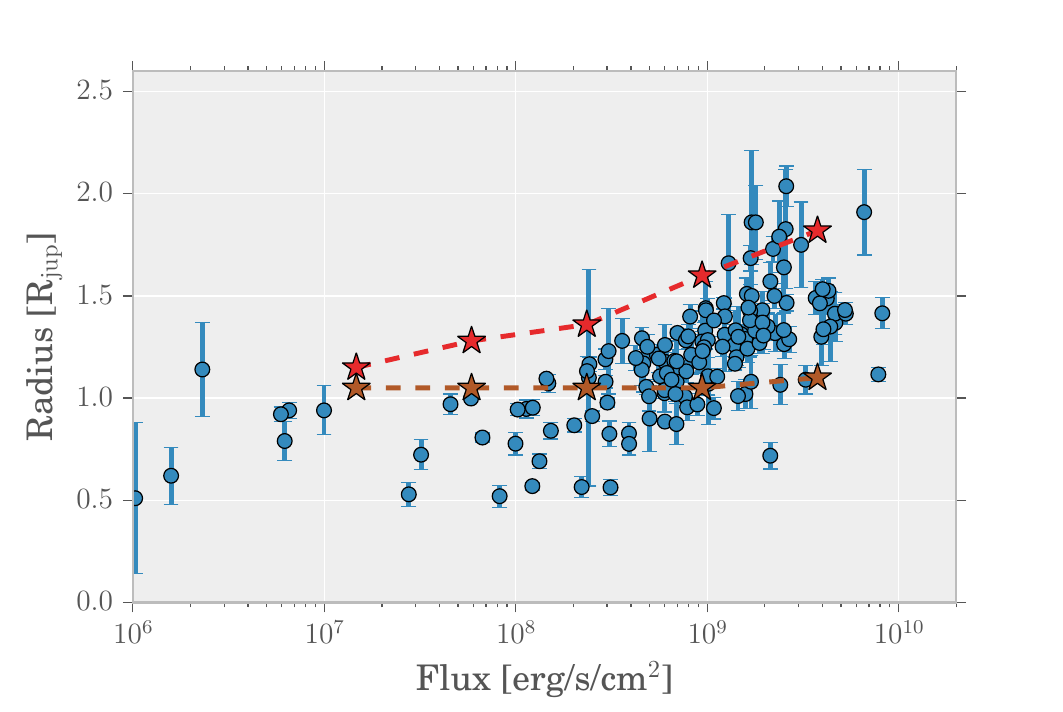}
\caption{\label{fig:rad_flux} Evolution of the radius as a function of irradiation for the 2D models (red) and for the 1D irradiated models (brown). All the atmospheric models are calculated with $\log(g)=2.97$. All the evolutionary models are calculated for an object mass 0.7 $M_\mathrm{jup}$ except for the most irradiated one for which we use a mass of 2 $M_\mathrm{jup}$ in order to keep $\log(g)=2.97\pm0.15$.}
\end{figure}

	There is now growing observational evidence to suggest a correlation between inflated radii and incident stellar flux \citep{Laughlin:2011hb,Miller:2011bq,Demory:2011ha,Weiss:2013kh,Figueira:2014hw}. \citet{Laughlin:2011hb} have found a best-fit dependence of $R_\mathrm{obs}-R_\mathrm{pred}\propto T_\mathrm{eff}^\beta$ with $\beta=1.4\pm0.6$ ($T_\mathrm{eff}$ is the equilibrium temperature of the planet defined as $T_\mathrm{star}\times(R_\mathrm{star}/a)^{1/2}$). In Fig.~\ref{fig:pt_profiles}, we show the PT profiles obtained with the 1D and 2D models as a function of irradiation flux.  All the other parameters are kept constant. We keep the same wind profile used for HD209458b (see Fig.~\ref{fig:inputs}), although the deep adiabat does not depend much on the wind magnitude (see Fig.~\ref{fig:cst_alpha_u} in appendix). It is clear that the stronger the irradiation, the hotter the internal adiabat the averaged 2D profiles connect with. We can easily estimate the magnitude of the temperature change at these levels ($\sim$200 bars) compared with the 1D models. For HD209458b (brown profiles), the profile becomes isothermal at about ~1800 K, $\sim$1 bar in 1D, whereas in 2D it becomes adiabatic at $\sim$5 bars with an adiabatic gradient of $(\gamma-1)/\gamma\approx$ 0.21 ($\gamma\approx1.27$). The temperature at 200 bars is thus now $1800\times(200/5)^{0.21}\approx 4000$ K.
Using evolutionary models \citep{Chabrier:2004db}, we can calculate the interior profiles, and the corresponding radii, that connect to the different atmospheric structures at the interior-atmosphere boundary. Fig.~\ref{fig:rad_flux} displays the observed correlation between these inflated radii and the irradiation flux (red lines), compared with the ones obtained in 1D (brown). The systems displayed in the plot are taken from \url{www.exoplanet.eu}. Note that we focus here on irradiated hot jupiters, i.e. objects with a dominant gaseous envelope with a lower limit radius of 1 R$_\mathrm{jup}$ in the absence of irradiation. Planets with smaller radii are less massive, e.g. hot Neptunes, and have a smaller gaseous atmosphere. Radius inflation for such objects is not an issue and can be easily explained by properly taking into account the irradiating flux \citep[e.g.][]{Baraffe:2008gl,Baraffe:2010fz}. For the sake of simplicity and to illustrate our purpose, we keep the same gravity in the atmospheric models, $\log(g)=2.97$, as a fiducial value, while varying the incident flux. In the evolutionary models, we use a mass of 0.7 $M_\mathrm{jup}$, the mass of HD209458b, except for the most irradiated object, for which we take a mass of 2 $M_\mathrm{jup}$, so all models are consistent with $\log(g)=2.97\pm0.15$. As shown in the figure (and in Tab.~\ref{tab:radius}) the 1D models hardly exceed 1.05 $R_\mathrm{jup}$. In contrast, the 2D models, all based on the same physical set-up reproduce the whole trend of increasing radius with increasing irradiation, including for the most inflated objects, with radii $\sim$2 $R_\mathrm{jup}$. Overall, the radii obtained in 2D seem to be a bit larger than the observed ones, a result for which we see different possible explanations:
\begin{itemize}
\item We used the same input profiles as the one obtained in 3D for HD209458b. Variations in the input parameters might yield slightly smaller radii. 
\item The steady-state 3D solutions might converge to intermediate profiles between the 1D and 2D ones. This is indeed suggested by our 3D simulation that seems to converge towards the 1.4 $R_\mathrm{jup}$ (the observed value) interior adiabat, whereas the 2D solution predicts a slightly larger value, 1.6 $R_\mathrm{jup}$.
\item Dissipation processes in the atmosphere (e.g. shear, ohmic dissipation that could reduce the magnitude of the vertical velocities) might lead to a slightly cooler adiabat and thus a smaller radius. 
\end{itemize}
Nevertheless, the 2D steady-state circulation models well reproduce the overall observed trend of enhanced radii of hot jupiters as a function of irradiation, based on a robust mechanism, namely atmospheric circulation induced by irradiation. The relation we obtain is $R_\mathrm{2D}-R_\mathrm{1D}\propto T_\mathrm{eff}^\beta$ with $\beta=1.43\pm0.14,$ in good agreement with the aforementioned one inferred from observations. 

\begin{table}
  \centering
  \begin{tabular}{l | r r}
    Flux & 1D radius & 2D radius  \\
    $[$erg/s/cm$^2]$ & [$R_\mathrm{jup}$] & [$R_\mathrm{jup}$] \\
    \hline
    \hline
    1.47$\times$10$^{7}$ & 1.05 & 1.15 \\
    5.88$\times$10$^{7}$ & 1.05 & 1.28 \\
    2.35$\times$10$^{8}$ & 1.05 & 1.36 \\
    9.41$\times$10$^{8}$ & 1.05 & 1.60 \\
    3.77$\times$10$^{9}$ & 1.1  & 1.82 
  \end{tabular}
  \caption{1D and 2D radii for a planet with $\log(g)=2.97$ for different irradiation fluxes.}
  \label{tab:radius}
\end{table}

\section{Discussion and conclusions}\label{sec:conclusion}

In this paper, we have presented a new approach to the problem of atmospheric circulation in tidally-locked, strongly irradiated exoplanets. We have constructed a 2D stationary circulation model that is able to predict the pressure/temperature and wind structure of the atmosphere at the equator, given a zonal wind profile at a given longitude and an estimate of the meridional to vertical mass flux ratio.
Since the 2D steady-state solution agrees well with the 3D simulation in the upper atmosphere, observational evidence of this mechanism remains relatively elusive at this stage. Nevetheless, it is worth mentionning that the steady-state solution does not have the isothermal plateau characteristic of the 1D solution nor does it present cold traps as in \citet{Parmentier:2016id}. The absence of these features might provide indirect observational constrains. 

With the increasing computational power of massively parallel architectures, a full 3D steady state might be reached sufficiently deep to better characterize the nature of the flow in the deep atmosphere. Nonetheless, the steady-state nature of our 2D circulation model provides a very useful complementary tool to current 3D simulations to understand the long-term physical processes at play in the atmospheres of hot jupiters. With this model, we have demonstrated that atmospheric circulation induced by irradiation can explain the inflated-radii of these planets. Indeed, the induced vertical mass flux and the advection of the potential temperature naturally constrain the atmosphere pressure/temperature profile to become adiabatic, even in the absence of convection, when the heating rate is small, as is the case in the deep atmosphere. Since this structure becomes adiabatic at smaller pressures than the one at which the atmosphere becomes unstable to convection, the 2D steady-state atmospheric profile reconnects with a hotter internal adiabat. This in turn implies a larger radius for a given mass, to explain the long-standing problem of the anomalously inflated radius of these objects.

\acknowledgments
Nathan Mayne acknowledges funding from the Leverhulme Trust via a Research Project Grant. James Manners acknowledges the support of a Met Office Academic Partnership secondment. This work is partly funded by the ERC grants No. 320478-TOFU and No. 247060-PEPS. Some of the calculations used the STFC DIRAC HPC service and the new University of Exeter Supercomputer ISCA.

\appendix
The stationary Euler equations in spherical coordinates $r$ (radius), $\phi$ (latitude), and $\lambda$ (longitude), supplemented by the energy equation for an ideal gas (whose internal energy e is given by the equation of state $P=(1-\gamma)e$) read

\begin{widetext}
\begin{align}
\frac{1}{r^2}\frac{\partial r^2\rho u_r}{\partial
  r}+\frac{1}{r\cos\phi}\frac{\partial \cos\phi\rho u_\phi}{\partial
  \phi}+\frac{1}{r\cos\phi}\frac{\partial\rho u_\lambda}{\partial
  \lambda}&=&0 \cr 
u_r \frac{\partial u_r}{\partial r}+\frac{u_\phi}{r}\frac{\partial
  u_r}{\partial \phi}+\frac{u_\lambda}{r\cos\phi}\frac{\partial
  u_r}{\partial \lambda} - \frac{u_\phi^2+u_\lambda^2}{r}
+\frac{1}{\rho} \frac{\partial p}{\partial r} + g(r)-2\Omega u_\lambda
\cos\phi  &=&0\cr
u_r\frac{\partial u_\phi}{\partial r}+\frac{u_\phi}{r}\frac{\partial
  u_\phi}{\partial \phi}+\frac{u_\lambda}{r\cos\phi}\frac{\partial
  u_\phi}{\partial \lambda}+\frac{u_r
  u_\phi}{r}+\frac{u_\lambda^2\tan\phi}{r} +\frac{1}{\rho
  r}\frac{\partial p}{\partial \phi}+2\Omega u_\lambda\sin\phi &=& 0\cr
u_r\frac{\partial u_\lambda}{\partial
  r}+\frac{u_\phi}{r}\frac{\partial u_\lambda}{\partial
  \phi}+\frac{u_\lambda}{r\cos\phi}\frac{\partial u_\lambda}{\partial
  \lambda}+\frac{u_r u_\lambda}{r}-\frac{u_\lambda u_\phi\tan\phi}{r} 
 +\frac{1}{\rho r\cos\phi}\frac{\partial p}{\partial
  \lambda}-2\Omega u_\phi\sin\phi+2\Omega u_r \cos\phi &=&0\cr
\frac{u_r}{\gamma-1}\frac{\partial p}{\partial
  r}+\frac{u_\phi}{\gamma-1}\frac{1}{r \cos\phi}\frac{\partial
  p}{\partial \phi}+\frac{u_\lambda}{\gamma-1}\frac{1}{r
  \cos\phi}\frac{\partial p}{\partial \lambda}+ 
\frac{\gamma p}{\gamma-1}\left(\frac{1}{r^2}\frac{\partial r^2
  u_r}{\partial r}+\frac{1}{r \cos\phi}\frac{\partial\cos\phi
  u_\phi}{\partial\phi} + \frac{1}{r \cos\phi}\frac{\partial
  u_\lambda}{\partial \lambda}\right) &=& H_\mathrm{rad} 
\end{align}
\end{widetext}

Where the velocity components are denoted by $u_r$, $u_\phi$ and $u_\lambda$, density is $\rho$, pressure $P$, $\Omega$ the angular velocity and $g(r)$ the gravitational acceleration. Since our radiative solver is plane-parallel, the heating rate at the equator is approximated as $H_\mathrm{rad}(r,\cos\phi,\cos\lambda)= -(\partial r^2 F_\mathrm{rad}/\partial r)/r^2 \approx -(\partial F_\mathrm{rad} (r,\cos\lambda))/\partial r$ for the conservation of energy. At the equator $\phi=0$, and assuming a constant residual meridional to vertical mass flux ratio $\alpha$:

\begin{equation}
\frac{1}{r^2}\frac{\partial r^2 \rho u_r}{\partial r} = \frac{1}{r\alpha} \frac{\partial
  \rho u_\phi}{\partial \phi}  
\end{equation}

We get the reduced system:

\begin{align}
  \frac{1}{r}\frac{\partial r^2 u_r}{\partial r}+r
  \frac{u_r}{\rho}\frac{\partial \rho}{\partial r} +\frac{\partial
    u_\phi}{\partial \phi} +\frac{\partial u_\lambda}{\partial
    \lambda} + \frac{u_\lambda}{\rho}\frac{\partial \rho}{\partial
    \lambda} &=& 0 \cr
  \frac{1}{r}\frac{\partial r^2 u_r}{\partial r} + r \frac{u_r}{\rho}\frac{\partial
    \rho}{\partial r} - \frac{1}{\alpha} \frac{\partial
    u_\phi}{\partial \phi} &=& 0 \cr
r u_r\frac{\partial u_\lambda}{\partial r}+ u_\lambda\frac{\partial
  u_\lambda}{\partial \lambda}+ u_r u_\lambda+ 2\Omega r u_r
+\frac{1}{\rho}\frac{\partial p}{\partial \lambda} &=&0 \cr
u_r\frac{\partial u_r}{\partial r} + \frac{u_\lambda}{r}
\frac{\partial u_r}{\partial \lambda}- \frac{u_\lambda^2}{r} -2\Omega
u_\lambda +\frac{1}{\rho} \frac{\partial p}{\partial r} + g(r) &=& 0 \cr
\frac{ru_r }{\gamma-1}\frac{\partial p}{\partial r} + \frac{u_\lambda
}{\gamma-1}\frac{\partial p}{\partial\lambda} &-& \cr 
 \frac{\gamma p}{\gamma-1}\left(r
  \frac{u_r}{\rho}\frac{\partial \rho}{\partial r} +
  \frac{u_\lambda}{\rho}\frac{\partial \rho}{\partial 
    \lambda} \right) + r\frac{\partial F_\mathrm{rad}}{\partial r} &=& 0
\end{align}

All the details of our radiative scheme with irradiation and isotropic scattering are given in \citet{Drummond:2016jg}. For the advection terms of the type $u_x \partial(⋅)/\partial x$, we used an upwind scheme. For stability reasons, we also needed to use an upwind (relative to the radiative flux transport, i.e. from bottom to top in the atmosphere) heating rate of the type $H_\mathrm{idepth}=-(F_\mathrm{idepth+1}-F_\mathrm{idepth})/(r_\mathrm{idepth+1}-r_\mathrm{idepth})$.
For axisymmetric irradiation (or absence of irradiation), an axisymmetric solution exists with $\partial u_\phi/\partial \phi=0$, $u_r=0$, $u_\lambda$ constant, and any 1D solution that fulfills the hydrostatic balance and the constant flux transport. Then, the system as proposed is not able to fully specify the wind solution, since any constant zonal wind is a solution. Therefore, we impose a zonal wind profile at a given longitude of the equator (the substellar point) which is equivalent to impose the total momentum in the zonal wind. 

\begin{figure*}[t]
\centering
\includegraphics[width=0.49\linewidth]{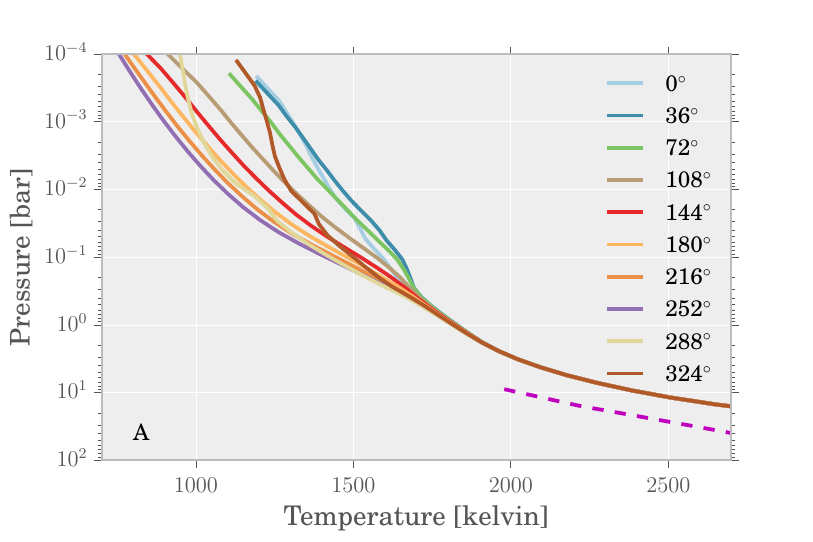}
\includegraphics[width=0.49\linewidth]{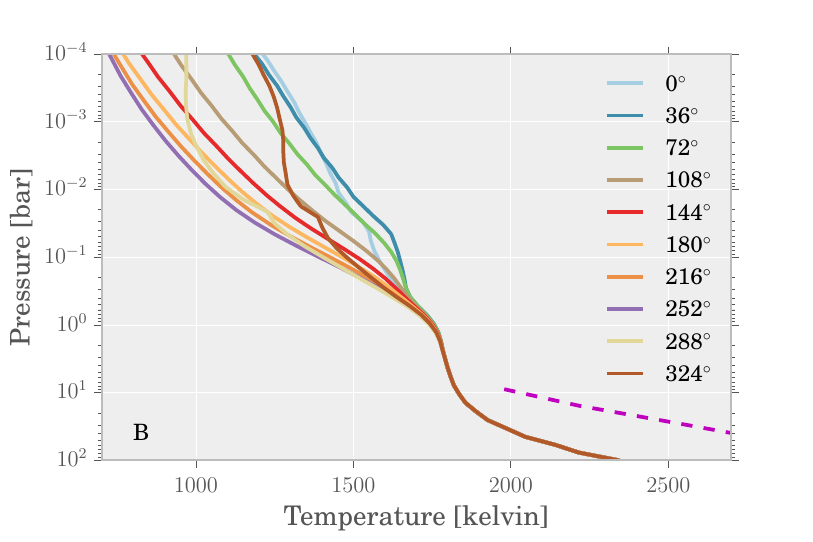}
\includegraphics[width=0.49\linewidth]{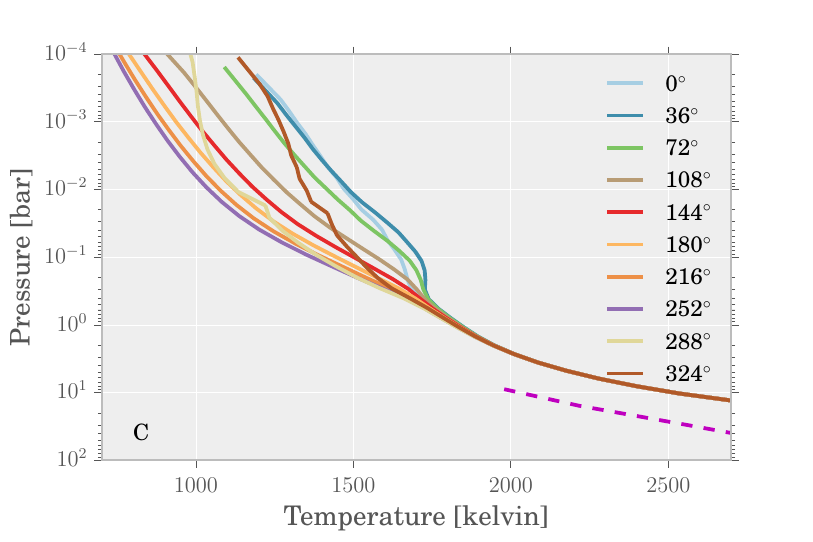}
\includegraphics[width=0.49\linewidth]{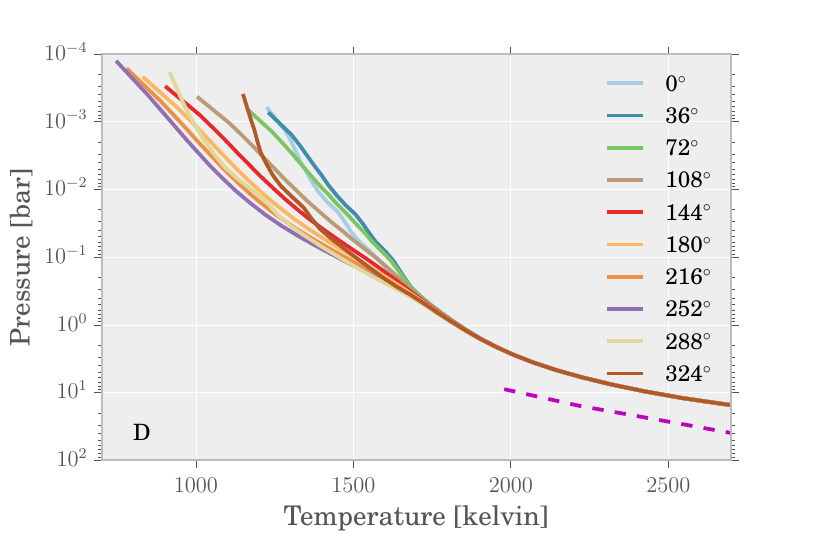}
\caption{\label{fig:cst_alpha_u} Pressure temperature profiles of the 2D circulation model with constant profiles $\alpha = 2$ and $u_\lambda=6$ km/s (A), $\alpha = 100$ and $u_\lambda=6$ km/s (B), $\alpha = 2$ and $u_\lambda=2$ km/s (C), $\alpha = 2$ and $u_\lambda=8$ km/s (D). The magenta dashed line is the same as in Fig.~\ref{fig:profiles}.}
\end{figure*}

In Fig.~\ref{fig:cst_alpha_u}, we show the 2D PT structures with constant profiles at $\alpha = 2$ and $u_\lambda=6$ km/s (A), $\alpha = 100$ and $u_\lambda=6$ km/s (B), $\alpha = 2$ and $u_\lambda=2$ km/s (C), $\alpha = 2$ and $u_\lambda=8$ km/s (D). With $\alpha=100$ throughout the whole atmosphere, we recover the usual isothermal plateau (at $T\approx 1800$ K), similar to the 1D solutions \citep[e.g.][]{Drummond:2016jg}. The occurence of an isothermal plateau is indeed expected, as $\alpha=100$ corresponds to a case for which the mass flux would be essentially purely meridional, with a negligible vertical contribution, a rather unlikely solution. Indeed, as seen in Fig.~\ref{fig:inputs}, the 3D run has already converged to an adiabatic structure at higher temperatures ($T\approx 2000$ K), for a pressure $P$ $\leq$ 10 bars. This suggests that the full 3D solution converges to the same adiabat as the 2D steady-state circulation model. The PT structures with different wind velocities at the substellar point show that the deep adiabat is only weakly dependent on this parameter. 

\bibliographystyle{aasjournal}
\bibliography{main.bib}

\end{document}